\documentclass{article}

\usepackage[total={6in, 8in}]{geometry}
\usepackage{authblk}
\usepackage{graphicx}
\usepackage{amssymb}
\usepackage{amsmath}
\usepackage{color}
\usepackage{url}
\usepackage{hyperref}
\usepackage{subcaption}
\usepackage{float}

\title{A data driven approach to classify descriptors based on their efficiency in translating noisy trajectories into physically-relevant information}

\author[1]{Simone Martino}
\author[1]{Domiziano Doria}
\author[1]{Chiara Lionello}
\author[1]{Matteo Becchi}
\author[1]{Giovanni M. Pavan\thanks{To whom correspondence should be addressed. E-mail: giovanni.pavan@polito.it}}

\affil[1]{Department of Applied Science and Technology, Politecnico di Torino, Torino 10129, Italy}

\date{\today}
\usepackage[style=numeric, maxnames=1, minnames=1, sorting=none]{biblatex} 
\begin{document}

\maketitle

\begin{abstract}
Reconstructing the physical complexity of many-body dynamical systems can be challenging. Starting from the trajectories of their constitutive units (raw data), typical approaches require selecting appropriate descriptors to convert them into time-series, which are then analyzed to extract interpretable information. However, identifying the most effective descriptor is often non-trivial. Here, we report a data-driven approach to compare the efficiency of various descriptors in extracting information from noisy trajectories and translating it into physically relevant insights. As a prototypical system with non-trivial internal complexity, we analyze molecular dynamics trajectories of an atomistic system where ice and water coexist in equilibrium near the solid/liquid transition temperature. We compare general and specific descriptors often used in aqueous systems: number of neighbors, molecular velocities, Smooth Overlap of Atomic Positions (SOAP), Local Environments and Neighbors Shuffling (LENS), Orientational Tetrahedral Order, and distance from the fifth neighbor ($d_5$). Using Onion Clustering -- an efficient unsupervised method for single-point time-series analysis -- we assess the maximum extractable information for each descriptor and rank them via a high-dimensional metric. Our results show that advanced descriptors like SOAP and LENS outperform classical ones due to higher signal-to-noise ratios. Nonetheless, even simple descriptors can rival or exceed advanced ones after local signal denoising. For example, $d_5$, initially among the weakest, becomes the most effective at resolving the system's non-local dynamical complexity after denoising. This work highlights the critical role of noise in information extraction from molecular trajectories and offers a data-driven approach to identify optimal descriptors for systems with characteristic internal complexity.
\end{abstract}

\newpage

\section*{Introduction}
The study of complex molecular systems composed of numerous constituent units, especially under conditions where they exhibit non-trivial internal dynamics, can be quite challenging. In such ensembles, interacting molecules can occupy a wide range of microstates, constantly exchanging among them. Untangling the complexity of such dynamical network is crucial for understanding the underlying physics of these systems, but it is often a difficult task~\cite{wolde_enhancement_1997, baletto_structural_2019, bochicchio_how_2019, gasparotto_identifying_2020, de_marco_controlling_2021, cioni_innate_2023}. Experimentally, it is typically challenging to obtain microscopic-level information with the spatial and temporal resolution necessary to discriminate and comprehend the processes occurring within these systems. In contrast, computational approaches such as Molecular Dynamics (MD) simulations can generate detailed trajectories for a given system, providing information on the individual trajectories of the units composing it, their mutual arrangements, and more. However, the extraction of meaningful information from these raw and often noisy datasets is often non-trivial. 

Using collective variables, or descriptors, can help to extract and retain key information from the raw MD trajectories, making them interpretable and useful for describing the phenomena and processes that characterize the system~\cite{kathirgamanathan_feature_2021, musil_physics-inspired_2021, schmidt_human-based_2023}. Typical approaches often study the behavior of a system by relying on intuitive, physically meaningful descriptors, such as distances, number of neighbors, and geometrical shape orders~\cite{nayar_comparison_2011}. However, a potential drawback of these approaches is that relying on descriptors based on human intuition can lead to biased or incomplete analysis, revealing only what was initially expected and potentially overlooking other important information. Another limitation of human-based descriptors is that they are often not transferable between different systems, which makes it more challenging to draw meaningful comparisons. 

To address these limitations, recent efforts have focused on developing more abstract and ``agnostic" descriptors, which do not rely on preconceived knowledge or physical intuition about the system~\cite{uhrin_through_2021, donkor_machine-learning_2023}. Some of these descriptors, based on particle density expansion, efficiently capture information about the local order and disorder in the positions of neighboring atoms (or molecules/units) in a given system. Notable examples include the Smooth Overlap of Atomic Positions (SOAP)~\cite{bartok_representing_2013}, the Atomic Cluster Expansion (ACE)~\cite{drautz_atomic_2019} or the $N$-body iterative contraction of equivariants (NICE)~\cite{nigam_recursive_2020} frameworks. 

Alternatively, there are descriptors of a different nature that do not quantify static features at each time step, but rather track how local features evolve over time -- the so-called dynamic descriptors. A relevant example of such abstract dynamic descriptor is the Local Environments and Neighbors Shuffling (LENS)~\cite{crippa_detecting_2023}, which tracks how the identity of the particles in the local environment surrounding each particle changes over time. LENS quantifies the dynamics of local environments, capturing significant fluctuations and providing insights into the system’s microscopic dynamical homogeneity or inhomogeneity~\cite{crippa_detecting_2023, becchi_layer-by-layer_2024, caruso_classification_2024, perrone_unsupervised_2024}. Another relevant example is {\it Time}SOAP ($t$SOAP)~\cite{caruso_timesoap_2023}, a one-dimensional quantity that tracks changes in the SOAP spectra of the particles along the trajectory. While LENS depends on the identities of neighboring particles without considering spatial coordinates, $t$SOAP is sensitive to their positions, so that the two descriptors capture different physical aspects of the particles' environment~\cite{crippa_machine_2023, caruso_classification_2024}. 

Aside from the ability of a specific descriptor to extract relevant information for a given system, another important point is finding the best approach to analyze the dataset once all the data are collected. Recent studies have shown, for example, that performing a global pattern recognition analysis on such datasets in the attempt of identify the microscopic environments within a system may only uncover statistically dominant patterns, often overlooking crucial information (such as their early emergence, or the presence of other less statistically relevant domains)~\cite{crippa_machine_2023}. Information loss can occur when time correlations within the time-series are neglected or when sparsely populated domains are masked by noise from dominant ones. Thus, incorporating time correlations into the analysis of each particle’s signal can reveal events that might otherwise be missed by traditional pattern recognition approaches~\cite{becchi_layer-by-layer_2024, butler_change_2024}. 

Given the large variety and diversity of descriptors that can be used to translate raw data (MD trajectories) into a dataset to analyze, a key question is which one is the best suited for a given system. In general, identifying the most appropriate descriptor is a crucial step, as preconceived choices -- often guided by prior experiences or previous studies -- may compromise or bias the final interpretation of the results. To address these important point, we present a purely data-driven and agnostic approach to effectively compare the efficiency of different descriptors in extracting and resolving the information contained in a trajectory. As a prototypical test case, we use a molecular trajectory with known non-trivial internal dynamical complexity -- specifically, a water system at the solid/liquid coexistence point~\cite{caruso_timesoap_2023, crippa_detecting_2023}. 

This system comprises various environments (solid ice, liquid water, liquid-solid interface, etc.) that are highly heterogeneous both structurally and dynamically, making it an ideal case study for assessing the effectiveness of local descriptors in resolving the system's complexity. We compare different types of static and dynamic descriptors, ranging from simpler human-based and physics-inspired ones to more abstract and data-driven ones that are general and agnostic. By employing a purely data-driven metric, we can quantify and compare the information extracted from the MD trajectory by each descriptor. This allows us to classify them based on their similarities and differences, as well as their efficiency in extracting and classifying information from the analyzed trajectories. Since a descriptor efficiency is determined by its signal-to-noise ratio, we also assess the impact of noise reduction across all explored descriptors using a recently reported approach~\cite{donkor_beyond_2024}. Our results show that even the simplest descriptors, once denoised, can be as efficient as the most advanced ones. This work highlights that it is more appropriate to discuss the best analysis framework rather than the best descriptor for extracting information from a specific system. Additionally, it provides a general, agnostic, physically interpretable, and data-driven approach to identify this framework based on a maximum resolved information criterion. 

\section*{Results}
\subsection*{Extracting information from trajectories}
\label{subs:extr_from_trjs}

\begin{figure}[H]
    \includegraphics[width=\textwidth]{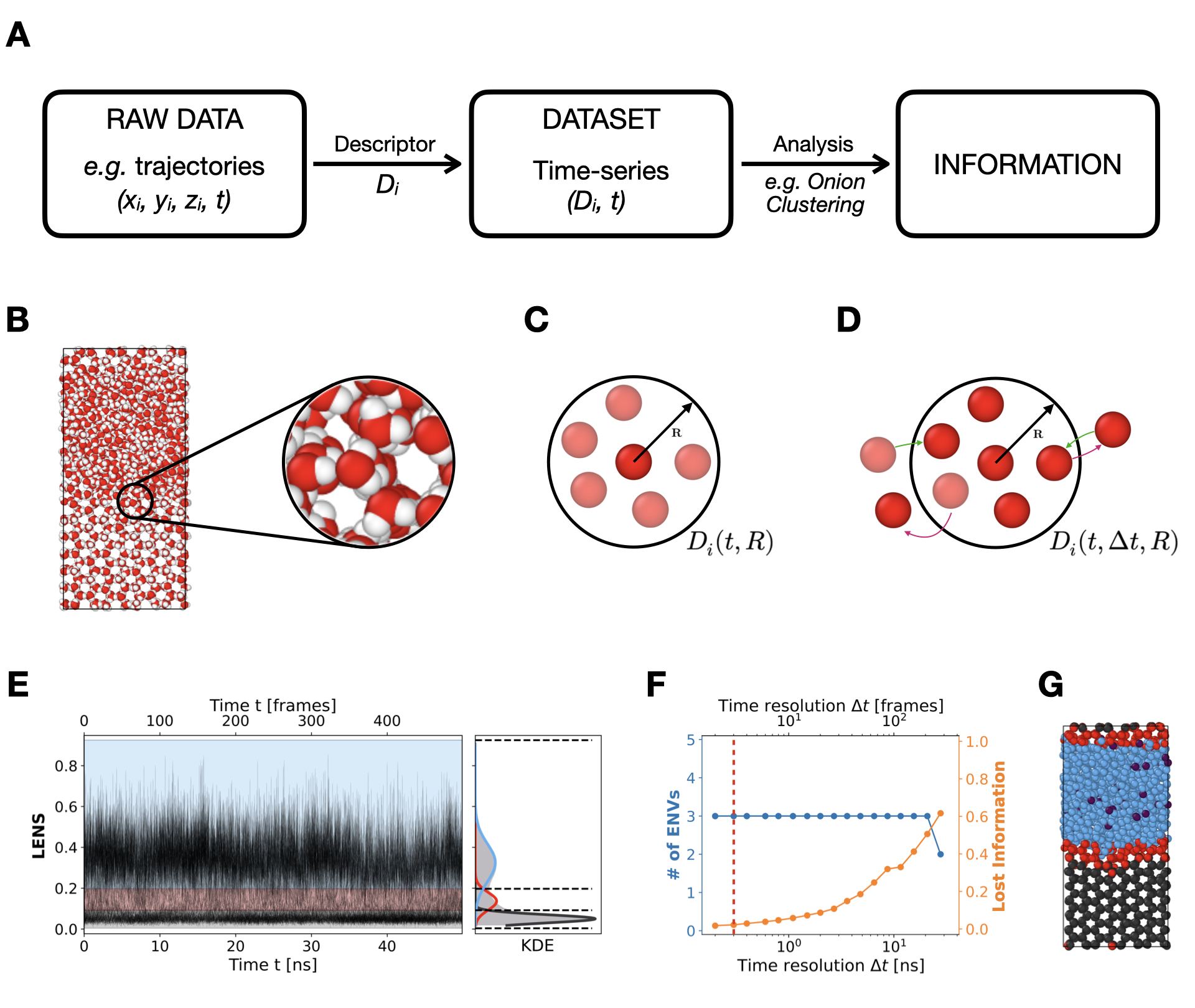}
    \caption{{\bf Ice/liquid water coexistence MD simulation, descriptors and LENS clustering}. {\bf A}: Flowchart of the information extracting procedure through the use of a generic descriptor $D_i$. 
    {\bf B}: Snapshot of the simulation. Oxygen atoms are colored red, and hydrogen atoms are colored white. 
    {\bf C}: Schematic representation of a static descriptor, $D_i(t, R)$, which depends on the coordinates and/or identities of molecules within a cutoff radius $R$ at time $t$. 
    {\bf D}: Schematic representation of a dynamic descriptor (LENS), which depends on the variation in the molecules' identities within a cutoff radius $R$ at times $t$ and $t+\delta t$. 
    {\bf E}: LENS signal time-series, for each particle as a function of simulation time. The background is colored according to the thresholds between the three identified clusters: solid ice (gray), liquid water (blue) and interface (red). The KDE of the signals (gray shaded area) is overlaid with the Gaussian distributions fitted by the clustering algorithm. Dashed lines represent the thresholds between the clusters. 
    {\bf F}: Typical Onion Clustering output plot, showing the number of environments detected (blue line) and the fraction of unclassified data points (orange line), as a function of the time resolution $\Delta t$. The red dashed lines indicates the time resolution ($\Delta t = 0.3$~ns) used for the clustering shown in this figure. 
    {\bf G}: Simulation snapshot where molecules are colored according to their cluster assignment (for $\Delta t = 0.3$~ns). Unclassified molecules are colored purple. }
    \label{fig1}
\end{figure}

Extracting information from a trajectory primarily involves two main steps. First, starting with the raw trajectory (the collection of $\Vec{r}_i(t)$ coordinates for every unit $i$ at every time-step $t$), the data are converted in a set of time-series $D_i(t)$ by selecting an appropriate descriptor $D$. Second, these time-series $D_i(t)$ are analyzed in various ways to extract meaningful information (Fig.~\ref{fig1}A). This general approach applies to the analysis of any trajectory, not limited to those obtained from simulations; here, as a prototypical example we consider a 50~ns-long MD trajectory of an atomistic model system composed of 2048 TIP4P/ICE molecules~\cite{abascal_potential_2005} in a rectangular simulation box, as shown in Fig.~\ref{fig1}B. The system is simulated with periodic boundary conditions, starting from a configuration in which 50\% of the molecules are in the liquid state and the remaining 50\% is in the solid state, arranged in a hexagonal ice crystalline structure ({\it Ih})~\cite{abascal_potential_2005}. After equilibrating the system at the melting temperature, we conducted a $50$~ns production run, saving the molecules' coordinates every $\delta t = 0.1$~ns (additional simulation details can be found in the methods section~\ref{subs:meth_simul}). 

Various descriptors can be used to extract information from these raw MD trajectories, broadly distinct between static (Fig.~\ref{fig1}C) and dynamic (Fig.~\ref{fig1}D). Static descriptors provide characteristic fingerprints of each molecule's local environment at each time-step; they are functions of the molecular coordinates at a specific time frame: $D_i(t) = D(\Vec{r}(t))$. Common descriptors used to study aqueous systems include the number of neighbors ($N_\text{neigh}$), the distance from the fifth neighbor ($d_5$), the orientational tetrahedral order parameter ($q_\text{tet}$), and the previously mentioned SOAP. 
In contrast, dynamic descriptors depend on the change in molecular environment between consecutive time-steps: $D_i(t) = D(\Vec{r}(t), \Vec{r}(t + \delta t))$. An example is LENS, which captures the local dynamics and diffusivity of each molecule’s environment by assessing the reshuffling and exchange among neighboring molecules within a certain cutoff distance~\cite{crippa_detecting_2023}. LENS primarily detects two types of changes in local environments over time: fluctuations in the number of neighbors (addition or departure of neighbors within $\delta t$) and changes in neighbor identity (swapping of molecules inside the cutoff with others outside within $\delta t$). This enables LENS to identify dynamically diverse microscopic (local) environments in a system, along with the fluctuations between them. 

Fig.~\ref{fig1}E depicts the LENS time-series computed for all water molecules in the system. The cumulative Kernel Density Estimation (KDE) of the LENS data points reveals two density peaks: one at LENS~$\sim0.1$, indicating a more static environment (corresponding to solid ice), and another at LENS~$\sim0.4$, identifying molecules in a more dynamic environment (corresponding to liquid water). However, there is more information in these data beyond the high-density peaks of the cumulative KDE. Analyzing these time series in detail requires distinguishing meaningful fluctuations from noise. Typical approaches involve single-point clustering of the time-series, which are examined over time rather than as a set of uncorrelated frames. Single point time-series clustering is a class of methods for grouping individual time-series data points based on their similarity in features, patterns, or dynamics. This approach allows for identifying and categorizing distinct temporal behaviors within a complex dataset, revealing underlying structures or states in the system~\cite{aghabozorgi_time-series_2015, aminikhanghahi_survey_2017, butler_change_2024, shiraj_anomaly_2024}. 

Among these methods, we chose a recently developed unsupervised clustering method, Onion Clustering, which ``peels" the dynamic complexity of noisy time-series, revealing all the classifiable information they contain. A relevant feature of Onion Clustering is that it requires the choice of a time resolution $\Delta t$, representing the minimum lifetime for an environment to be considered stable. The key factor we highlight here is that, instead of relying on a specific {\it a priori} choice of $\Delta t$ (which can lead to a potentially risky black-box approach), Onion Clustering automatically performs single-point clustering across all possible $\Delta t$ values -- from the highest possible resolution ($\Delta t = 2$~frames) to the lowest (the entire trajectory length, $\Delta t = 500$~frames for the current simulation). For each $\Delta t$, it outputs both the number of clusters that can be statistically robustly distinguished at that resolution and the fraction that cannot be classified due to insufficient resolution (indicating dynamical events occurring faster than $\Delta t$). For a detailed description of Onion Clustering, we refer readers to~\cite{becchi_layer-by-layer_2024}. 

Fig.~\ref{fig1}F shows the results of Onion Clustering for different $\Delta t$ values in the analysis of the LENS time-series extracted for this system under study. The number of classifiable clusters at each $\Delta t$ is shown in blue, while the fraction of  unclassified data points (lost information) is colored in orange. The Onion results demonstrate that, down to a resolution of $\Delta t < 20$~ns, three statistically-relevant environments can be resolved in the LENS time-series. Fig.~\ref{fig1}G colors the water molecules based on their assigned clusters at the example resolution of $\Delta t = 0.3$~ns, clearly identifying three physically relevant and dynamically different clusters: solid ice (dark gray), the ice-water interface (red), and liquid water (blue). The small fraction of unclassified molecules is represented in purple, corresponding to molecules undergoing transitions faster than the selected time resolution of the analysis. Notably, beyond $\Delta t=20$~ns, the efficiency decreases, and the number of distinguishable clusters drops to 2, with the ability to distinguish the interface being lost. 
\subsection*{Comparison between different descriptors}
\begin{figure}[H]
    \includegraphics[width=\textwidth]{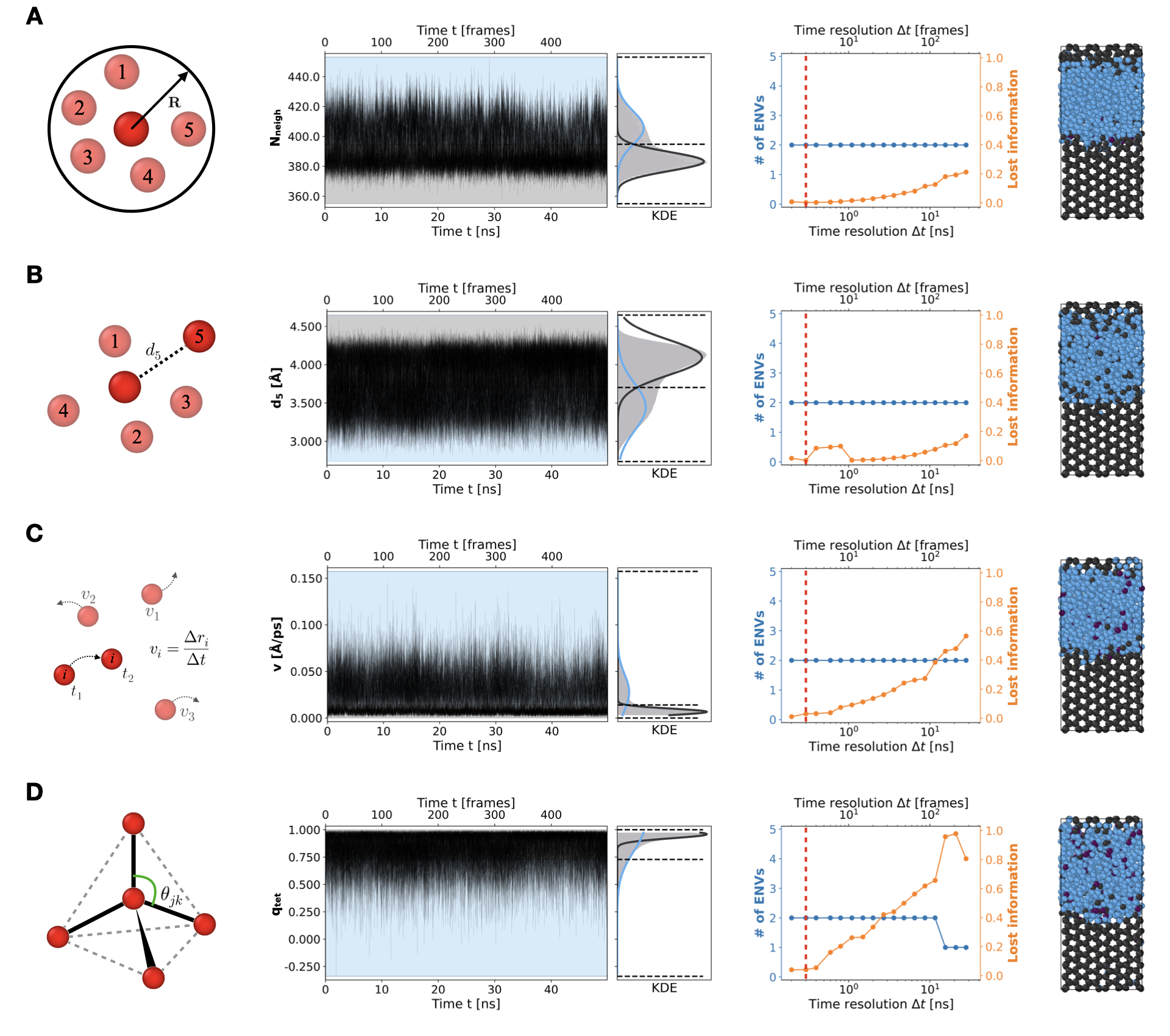}
    \caption{{\bf Comparison between descriptors.} For the different descriptors -- {\bf A}: number of neighbors $N_\text{neigh}$, {\bf B}: distance from the fifth atom $d_5$, {\bf C}: molecule velocity modulus $v$ and {\bf D}: orientational tetrahedral order parameter $q_\text{tet}$ -- we show, from left to right: a schematic representation, the signal distribution over the entire simulation, clustered KDE, the number of resolved environments (blue line) and fraction of unclassified data (orange line), and a clustering-colored snapshot in correspondence to the time resolution highlighted with the red line. }
    \label{fig2}
\end{figure}

In this study, we compared various descriptors commonly used in water simulations to assess their effectiveness in capturing the system's features. As examples of more general descriptors we selected the number of neighbors within a defined cutoff distance $N_\text{neigh}$ (Fig.~\ref{fig2}A), and the modulus of each molecule's velocity $v$ (Fig.~\ref{fig2}C). For descriptors specifically designed for aqueous systems, we considered the distance from the fifth neighbor $d_5$ (Fig.~\ref{fig2}B), and the orientational tetrahedral order parameter $q_\text{tet}$ (Fig.~\ref{fig2}D). More details on the computation of these descriptors can be found in Section~\ref{subs:meth_des}. 
Fig.~\ref{fig2}A-D displays the time-series generated from the MD trajectory for each of these physics-inspired descriptors, together with the clusters identified by Onion Clustering, shown with solid colored lines on the KDE of the signals. As can be seen from the plots and snapshots in Fig.~\ref{fig2}, for all the descriptors at most two different environments (solid ice and liquid water) are identified. With none of these descriptors the clustering algorithm is able to resolve the solid/liquid interface, differently from what we found using LENS. 
These results indicate that these descriptors are less effective than LENS in characterizing the system’s internal physics. Specifically, while they reliably distinguish between ice and liquid phases, they fail to capture the distinct structural and dynamical features of the interface region. This outcome suggests that descriptors specifically tailored for a particular system -- in this case, aqueous environments -- do not necessarily outperform more abstract, general descriptors. Furthermore, both static and dynamic descriptors yielded similar results, with no clear advantage in using one type over the other for identifying the key features in this system. 
\subsection*{A more advanced high-dimensional descriptor: SOAP}
\label{subs:SOAP}

One powerful and widely used descriptor in the study of complex molecular systems, including aqueous ones, is the Smooth Overlap of Atomic Position (SOAP). SOAP provides a rotational-invariant decomposition of the local particle density of the neighborhood around each particle within a defined cutoff radius~\cite{bartok_representing_2013}. For each particle $i$ at each simulation frame $t$, SOAP generates a spectrum of coefficients that encode a specific fingerprint of the relative positions of neighboring particles. 

Recently, it has been shown~\cite{donkor_beyond_2024} that in systems with high structural heterogeneity (for example, supercooled water below the liquid-liquid phase transition~\cite{harrington_liquid-liquid_1997}), the ability of SOAP to identify different molecular environments can be greatly improved by local noise reduction. SOAP spectra are often ``overloaded" with local structural information, as well as noise from surrounding environments. This issue can be mitigated by averaging the SOAP vectors of molecule $i$ with those of its neighbors within the cutoff sphere. By doing so, the local heterogeneity within the microscopic environment surrounding each particle is smoothed, reducing noise and enhancing SOAP's ability to distinguish between different environments, particularly those with non-local differences. This noise reduction approach has recently been shown to be highly efficient, enabling, for example, the distinction of the presence and coexistence of two liquid phases in aqueous systems -- an otherwise challenging task. 

\begin{figure}[H]
    \includegraphics[width=\textwidth]{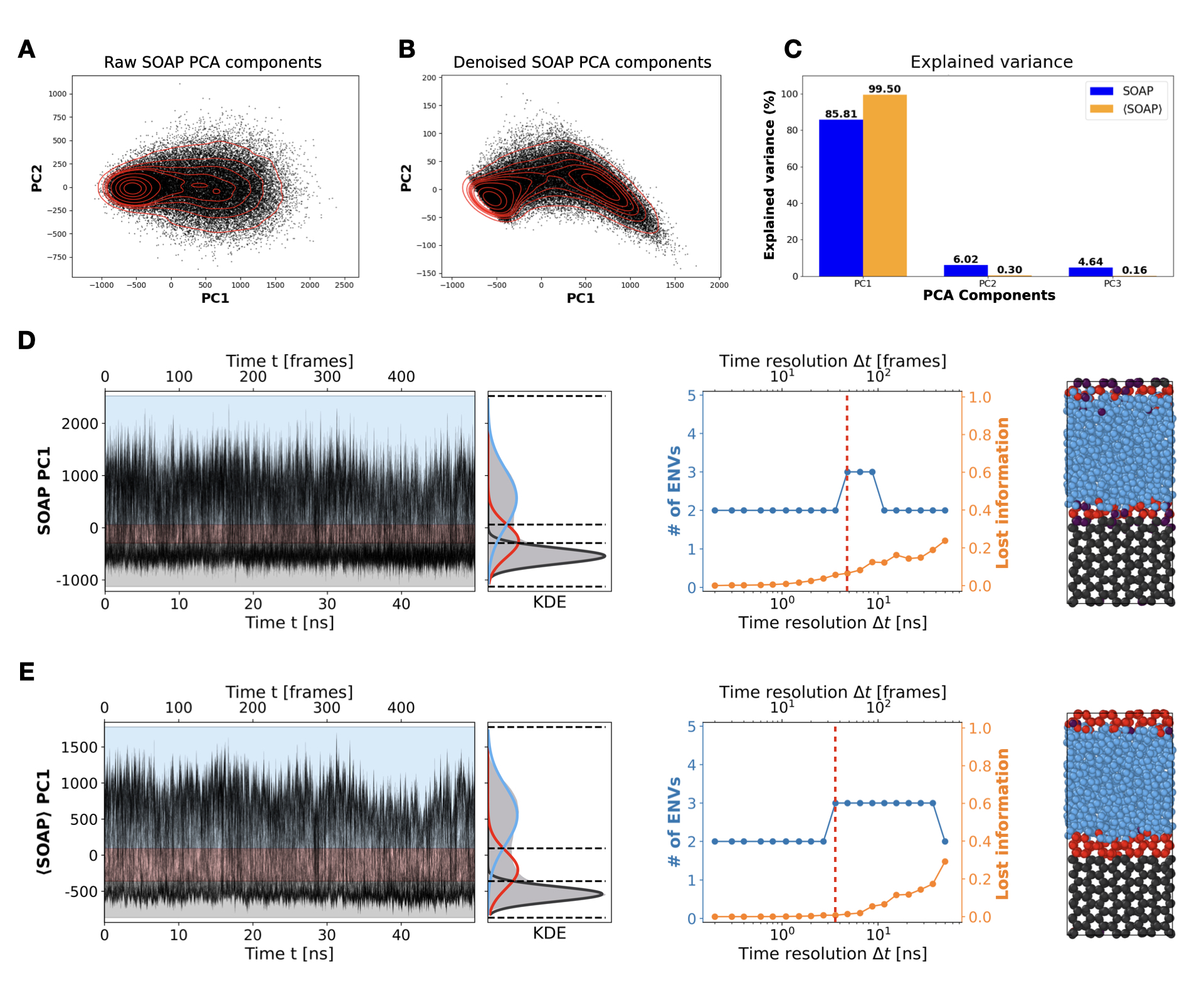}
        \caption{{\bf Clustering results on SOAP}. {\bf A}: Projection of SOAP on the first two PCs. {\bf B}: Projection of spatially averaged SOAP over the first two PCs. {\bf C}: First three PCs' explained variance of raw SOAP (blue bars) and spatially averaged (orange bars). {\bf D}: Onion Clustering output on SOAP; from left to right, signal distribution, clustered KDE, number of resolved environments (blue line), fraction of unclassified data (orange line) and clustering-colored snapshot in correspondence to the time resolution highlighted with the red line. {\bf E}: Onion Clustering output on spatially averaged SOAP; same as in panel {\bf D}. }
    \label{fig3}
\end{figure}

Fig.~\ref{fig3}B shows the PCA projection of the SOAP spectra of water molecules after smoothing. Compared to the raw SOAP dataset in Fig.~\ref{fig3}A, it is clear that smoothing the local noise facilitates the detection of the liquid domain in our water-ice system. Fig.~\ref{fig3}C shows the variance explained by the first three principal components (PC). In the raw SOAP dataset, the sum of the first three principal components explains approximately $96,5$\% of the variance, which makes it a good approximation of the entire dataset. We spatially averaged (smoothed) the SOAP spectra of each molecule with those of the neighboring molecules within a cutoff radius of 10 \AA (the same used for the SOAP calculation). As a result, the first three components explains approximately $99.96$\% of the total variance (orange bars in~\ref{fig3}C), with PC1 alone accounting for about $99.5$\%. 

Recently, it has been shown \cite{lionello2024relevant} that performing a single-point time-series clustering (using, for example, Onion Clustering) on PC1 alone typically provides more insightful information than a pattern recognition analysis applied to the entire dataset. For instance, Fig.~\ref{fig3}D-E shows the results of Onion Clustering on the PC1 time-series computed for both the raw (Fig.~\ref{fig3}D) and spatially-smoothed (Fig.~\ref{fig3}E) SOAP datasets. Onion Clustering successfully identifies three distinct environments in these time-series: solid ice (dark gray), liquid water (blue), and the solid-liquid interface (red). Notably, the application of local noise reduction via spatial averaging increases the separation between the KDE peaks of these different environments, thus improving detection and classification. This effect is reflected in the PCA components, where local noise reduction boosts the explained variance in PC1 from approximately $85.8$\% to $99.5$\% in PC1 (Fig.~\ref{fig3}C). 
\newpage
Furthermore, regarding the dependency of the clustering results on the time resolution ($\Delta t$), Fig.~\ref{fig3}D-E shows that the interface is only detected within a specific range of $\Delta t$: $4.8<\Delta t<8.7$ ns in the raw trajectories, compared to $3.6<\Delta t<37.2$ ns in the denoised ones. The lower resolution limit is determined by the minimum observation time required to gather sufficient information to distinguish one environment from another. The upper resolution limit, which is determined by the average lifetime of the particles inside the interface, is also affected by the noise in the dataset. As the level of noise in the data increases, the overlap between the signals from different environments grows, making it more challenging to reliably classify longer segments of the trajectories into distinct environments. As shown in the plots, denoising the SOAP data shifts the upper limit to much higher values, improving the robustness of interface detection. This is due to the reduction in noise, which increases the signal-to-noise ratio in the smoothed PC1 SOAP time-series compared to the raw data. In the raw SOAP data, the noise from the two most populated environments (solid ice and liquid water) is so large that it becomes impossible to discern the interface signal for most choices of $\Delta t$. 

Since each different descriptor $D_i$ has its own signal-to-noise ratio, methods such as this one, based on spatial averaging, provide a viable strategy to create denoised versions of virtually any descriptor. This opens up the possibility to expand studies assessing the effect of noise on the efficiency of different descriptors in capturing the complex physics of systems like the one examined here. 
\subsection*{Cleaning descriptors from local noise}
\begin{figure}[H]
    \includegraphics[width=\textwidth]{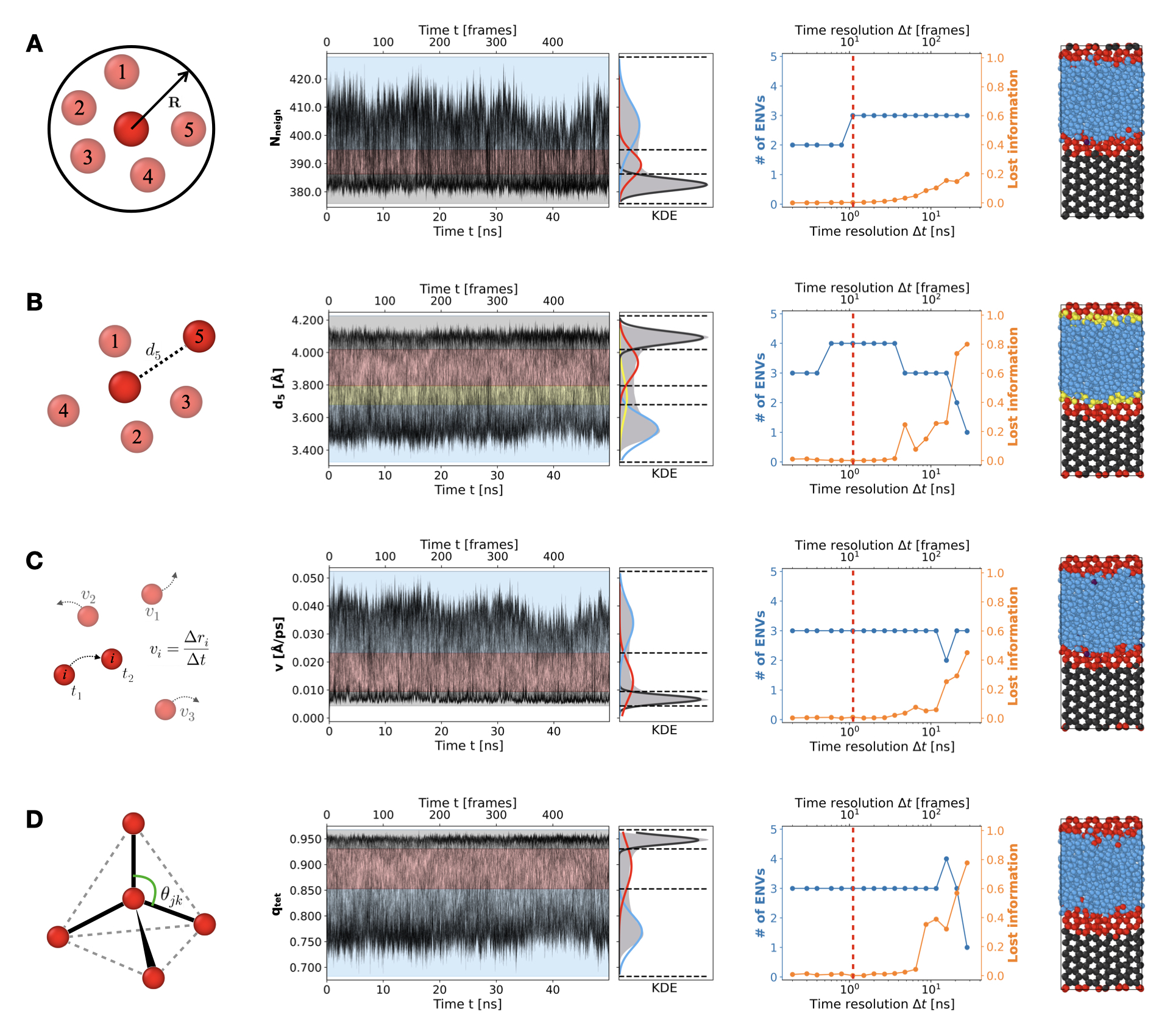}
    \caption{{\bf Comparison between descriptors after local noise reduction.} For each spatially averaged descriptor -- {\bf A}: number of neighbors $N_\text{neigh}$, {\bf B}: distance from the fifth atom $d_5$, {\bf C}: molecule velocity modulus $v$ and {\bf D}: orientational tetrahedral order parameter $q_\text{tet}$ -- we show, from left to right: a schematic representation, the signal distribution over the entire simulation, clustered KDE, the number of resolved environments (blue line) and fraction of unclassified data (orange line), and a clustering-colored snapshot in correspondence to the time resolution highlighted with the red line. }
    \label{fig4}
\end{figure}

The improvement observed in SOAP results suggests that a similar denoising approach could be applied to all previously discussed descriptors. The outcomes, displayed in Fig.~\ref{fig4}, reveal that spatial averaging significantly enhances the performance of all descriptors. After local noise reduction, descriptors such as $N_\text{neigh}$, $v$, and $q_\text{tet}$ become capable of detecting the interface as a distinct environment with efficiency comparable to (or exceeding) that of more sophisticated descriptors like LENS and SOAP. An especially intriguing result is obtained with $d_5$: it not only identifies the interface but also distinguishes between two subregions -- one exposed to liquid and the other to ice. 

Interestingly, spatial averaging yields a less pronounced benefit for LENS. In the raw data, LENS already stands out as a top-performing descriptor, with a distribution that is easily classified by Onion Clustering without further processing. Consequently, spatial averaging provides only a slight improvement, similar to what is observed for SOAP. Both LENS and SOAP possess a high intrinsic signal-to-noise ratio, and while spatial averaging does increase clarity, the impact is far more limited than with simpler descriptors. Denoised LENS results can be found in Fig.S1.

\subsection*{An ``evaluation space'' for comparing descriptors}
The application of Onion Clustering shows how different descriptors can resolve varying numbers of microstates depending on the time resolution. In the previous section, we saw how optimizing the signal-to-noise ratio of each time-series improves both the number and robustness of resolvable microstates. Building on this, we leverage the extensive information provided by Onion Clustering to construct an ``evaluation space" -- a framework where each tested descriptor can be positioned and quantitatively compared using a data-driven metric. Similar approaches have recently been applied to, for example, compare lipid bilayers modeled with different force fields~\cite{capelli_data-driven_2021} or classify different types of self-assembled soft materials by assessing the similarity of their local molecular (SOAP) environments~\cite{gardin_classifying_2022}. 

\begin{figure}[H]
    \includegraphics[width=\textwidth]{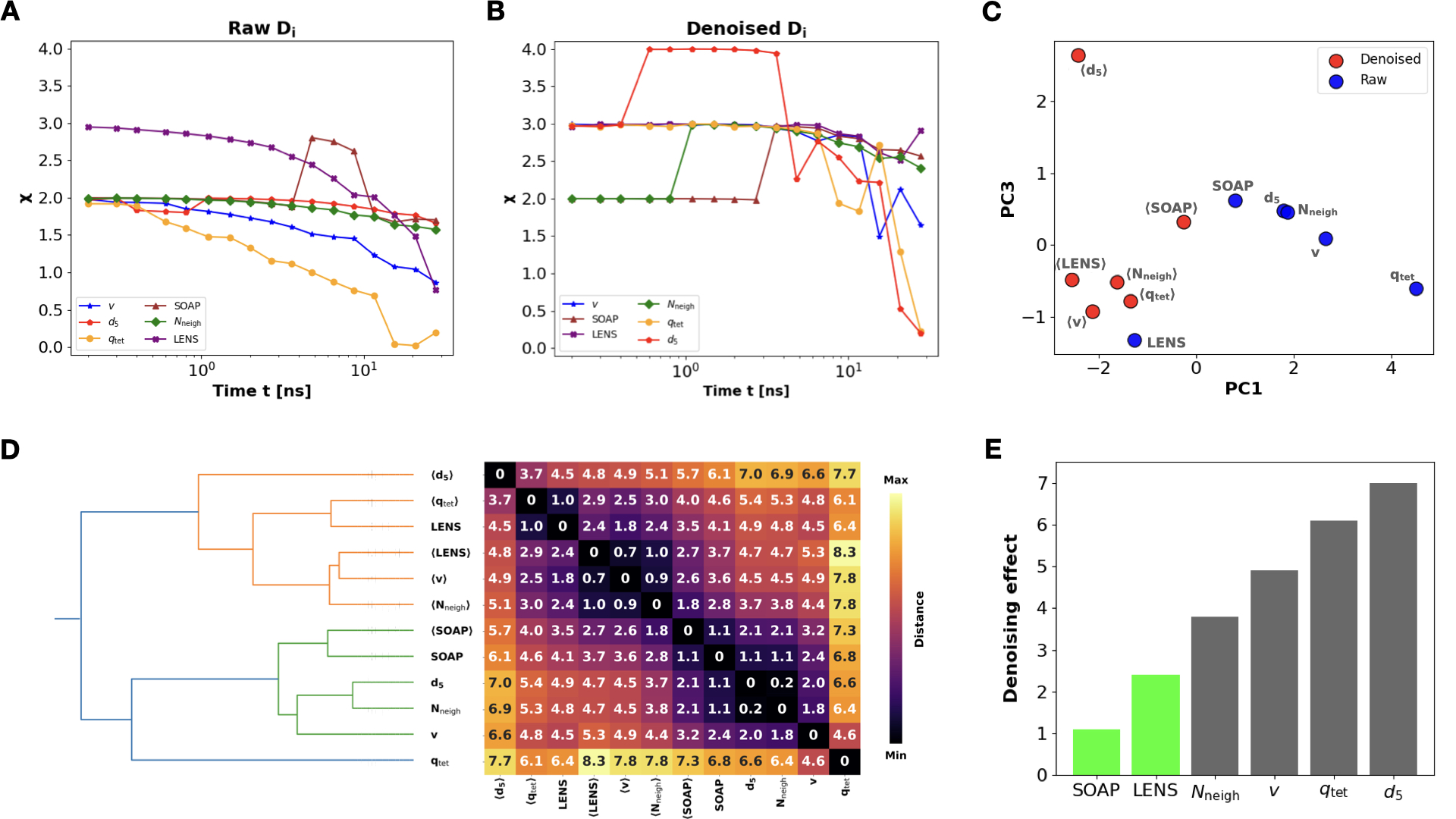}
    \caption{{\bf Evaluation space.} ({\bf A}) $\chi$ parameter for each raw descriptor, as a function of the time resolution $\Delta t$. ({\bf B}) $\chi$ parameter for each denoised descriptor, as a function of the time resolution $\Delta t$. ({\bf C}) Two dimensional projection of the ``evaluation space'', using the first and the third PCs. ({\bf D}) Hierarchical clustering results and distance matrix between all the descriptors studied. ({\bf E}) Amplitude of the denoising effect, obtained computing the distances between raw version of the descriptors and their respective denoised one.} 
    \label{fig5}
\end{figure}

The output of Onion Clustering can be summarized by the number of resolved environments $n_\text{env}$ and the fraction of information lost $f_0$ for any given time resolution $\Delta t$, which we combine into a single metric, $\chi$, defined as: 
$$ \chi(\Delta t) = n_{\text{env}}(\Delta t) \cdot\left[1-f_0(\Delta t)\right]$$
This metric $\chi$ captures the effectiveness of a descriptor in extracting meaningful information from noisy trajectories and in processing it into statistically distinct micro-environments. Fig.~\ref{fig5}A-B shows the $\chi$ values across different time resolutions and descriptors used. 

To build an ``evaluation dataset” capturing each descriptor's performance in resolving system complexity, we selected specific features output by Onion Clustering. We included seven key features: (i) the mode of the number of states identified, (ii) the number of occurrences of each state count, (iii) the time resolution at which more than 50\% of the information is lost, (iv) the mean fraction of information lost before it exceeds 50\%, (v) the standard deviation of the information lost fraction before it exceeds 50\%, (vi) the maximum $\chi$ value, and (vii) the average $\chi$ value before the information lost fraction surpasses 50\% (for a more detailed description of this dataset see SI). This process creates a multi-dimensional dataset which enables an in-depth comparison of descriptors, with flexibility to add further features as needed. 

To create the “evaluation space”, we performed a PCA dimensionality reduction of this dataset. Fig.~\ref{fig5}C shows the first and the third principal components, color-coded to distinguish between the raw (blue) and denoised (red) versions of each descriptor (the other PCs are available in Fig.~S2). This representation allows for a hierarchical clustering analysis~\cite{bar-joseph_fast_2001, mullner_modern_2011} based on distances between the PCA scores, enabling a comparative view of the descriptors (Fig.\ref{fig5}D). 

It is important to emphasize that this method does not directly rank descriptors from best to worst; rather, it defines a hierarchy of similarity and difference, showing which descriptors are closely related in terms of extracted information and which ones stand apart quantitatively. In this analysis, we observe a clear distinction between denoised and raw descriptors, except for the raw LENS descriptor, which already stands out in its ability to identify the water/ice interface. At the extremes of this evaluation space, denoised $d_5$ and raw $q_\text{tet}$ are the most distinctive descriptors, for contrasting reasons. Denoised $d_5$ uniquely captures the second interface across a significant range of time resolution, while $q_\text{tet}$ is a descriptor with low signal-to-noise ratio that struggles to capture relevant information across the whole time resolution range. 
\section*{Conclusions}
In the study of complex many-body systems, selecting an effective analysis framework and identifying the most informative descriptors are critical steps for extracting meaningful information from inherently noisy trajectories of individual components. While choosing the optimal descriptor for a given system might be challenging, in this work we introduced a data-driven, parameter-free approach to address this crucial aspect, which is fundamental to virtually all types of analyses in complex system research. 

To evaluate the effectiveness of different descriptors in capturing and resolving information, we exploited Onion Clustering, a single-point time-series clustering algorithm with two key advantages. Through an iterative find-classify-archive approach, Onion Clustering reveals all classifiable information based on the time resolution ($\Delta t$), thereby automatically identifying the optimal resolution for analysis. This also allows it to estimate the amount of data that cannot be classified in a statistically robust manner due to resolution limitations. By leveraging this information, we compared various descriptors in characterizing molecular environments in a water/ice coexistence system at the atomic level. 

Our results demonstrate that general-purpose descriptors, such as LENS and SOAP, are able to identify and classify physically relevant molecular microstates more effectively than descriptors specifically tailored for this system, thanks to their inherently higher signal-to-noise ratio. 
additionally, we show that performance across many descriptors can be sensibly enhanced through local denoising via spatial averaging. This approach boosts the signal-to-noise ratio in time-series data, enabling even simple descriptors like the number of neighbors $N_{\text{neigh}}$ to perform comparably to more advanced ones such as SOAP and LENS, after denoising. 

A quantitative comparisons of the descriptors was achieved by constructing an ``evaluation space" that enables the use of an Onion Clustering-based metric to quantify the similarities between descriptors in terms of their effectiveness in resolving noisy trajectories and extracting physically meaningful information. The method is completely parameter-free and data-driven, and it does not require arbitrary choices or prior knowledge when assessing performance. Additionally, the number of features included in the evaluation dataset can be expanded by incorporating additional parameters as needed. 

The results reported here also demonstrate the extent to which different descriptors can be improved through local denoising. This suggests that, for molecular systems with intricate internal structures, it is more productive to pursue a customized analysis framework than to rely on any single “best” descriptor. 

We believe this approach offers a valuable framework for optimizing the selection of descriptors and analysis methods in the study of complex dynamical systems. Its versatility extends beyond molecular systems, making it suitable for applications involving time-series data from mesoscopic and macroscopic systems, as well as noisy trajectory datasets obtained experimentally~\cite{keogh2002finding, gupta2013outlier, mantegna1999introduction, nagy_hierarchical_2010}. The method’s generality and data-driven nature thus provide a flexible tool for extracting meaningful insights across a wide range of scales and disciplines. 
\setcounter{section}{1}
{\small
\section*{Methods}
\subsection{Details on the water/ice simulation}
\label{subs:meth_simul}
The molecular ice/water system is simulated using the direct co-existence technique, with the TIP4P/Ice water model~\cite{abascal_potential_2005}. The trajectory is obtained starting from a configuration of 50\% ice 50\% liquid, at $T=268$ K. After equilibration, a production run is performed for $t = 50$~ns, sampled and analyzed every $0.1$~ns. GROMACS software is used to run the simulation~\cite{abraham_gromacs_2015}. More detailed information can be found in~\cite{crippa_detecting_2023}. 

\subsection{Details on the descriptors}
\label{subs:meth_des}

\paragraph{Smooth Overlap of Atomic Position (SOAP).} This descriptor~\cite{bartok_representing_2013} is a representation of the atomic neighbor density of a particle

$$\rho(\mathbf{r}) = \sum_{i=1}^{n_{cut}}\delta(\mathbf{r} - \mathbf{r}_i)$$
where $i$ runs over the $n_{cut}$ particles closer than a cutoff radius $r_{cut}$ from the central particle (in this work $r_{\text{cut}}=10$ \AA), and $\mathbf{r}_i$ is the position of the $i-$th particle. $\rho(\mathbf{r})$ can be approximated in terms of radial basis functions $g_n(r)$ and spherical harmonics $Y_{lm}(\theta, \phi)$ as

$$\rho(\mathbf{r}) \approx \sum_{n=0}^{n_\text{max}}\sum_{l=0}^{l_\text{max}}\sum_{m=-l}^{l}c_{nlm}\cdot g_n(r) \cdot Y_{lm}(\theta, \phi)$$
where $c_{nlm}$ are the expansion coefficients. In this work $n_{\text{max}}$ and $l_{\text{max}}$ have been set to 8. From this expansion, it is possible to compute a rotational-invariant power spectrum $\mathbf{p}$, whose components are

$$p_{nn'l}= \pi\sqrt{\frac{8}{2l + 1}}\sum_m(c_{nlm})^\dagger c_{n'lm}$$
which contains the structural information on the particle's environment. 
SOAP power spectra were computed using the DScribe~\cite{himanen_dscribe_2020, laakso_updates_2023} and Dynsight~\cite{noauthor_dynsight_nodate} packages. 

\paragraph{Local Environments and Neighbors Shuffling (LENS).}  The value of LENS between time $t$ and $t+\Delta t$ is defined as
$$
\delta_i(t) = \frac{\#(C_i^t \cup C_i^{t+\Delta t}) - \#(C_i^t \cap C_i^{t+\Delta t})}{\#(C_i^t) + \#(C_i^{t+\Delta t})}
$$
and represents the set difference between the mathematical union and intersection of neighbor IDs lists present within $r_{\text{cut}}$ ($10 \text{\AA}$ in this work) from particle $i$ at the two consecutive time steps $t$ and $t+\Delta t$,  normalized by the total length of the neighbor ID lists. Here, $C_i^t$ is the list of the IDs of the neighboring particles of particle $i$, and $\#(C_i^t)$ is its cardinality. The result is a value between 0 and 1, where 0 means that no changes happened in the neighbor list, while 1 means that all the neighbors' identities changed. Full details are available in~\cite{crippa_detecting_2023}. LENS was computed using the Dynsight~\cite{noauthor_dynsight_nodate} package. 

\paragraph{Orientational tetrahedral order parameter $q_\text{tet}$.} This quantity has been often used~\cite{chau_new_1998, giovambattista_structural_2005} to measures how much a local environment resembles a perfect tetrahedron; it is computed as 

$$q_\text{tet} = 1 - \frac{3}{8}\sum_{j=1}^3\sum_{k=j+1}^4\left(\cos\phi_{jk} + \frac{1}{3}\right)$$
where $\phi_{jk}$ is the angle between the central molecule and its neighbors $j$ and $k$. $q_\text{tet}$ equals 1 if the molecules form a perfect tetrahedron, while it's closer to 0 for non-tetrahedral environments. 

\paragraph{Other descriptors.} The use of  distance from the fifth neighbor ($d_5$) descriptor to characterize different environments in water is shown, for instance, in~\cite{cuthbertson_mixturelike_2011, donkor_machine-learning_2023}. Number of neighbors ($N_\text{neigh}$) and the modulus of the instantaneous particles velocity ($v$) are widely used general descriptors in various field of application~\cite{donkor_machine-learning_2023}. With regards to the number of neighbors, in this work $R = 10$ \AA~is used. 

\subsection{The Onion Clustering algorithm}
\label{subs:meth_onion}
Onion Clustering is an algorithm for single-point clustering of time-series data. It performs a series of clustering analyses, each one with a different time-resolution $\Delta t$, which is the minimum lifetime required for a cluster to be characterized as a stable environment. The clustering proceeds in an iterative way. At each iteration, the maximum of the cumulative distribution of data points is identified as a Gaussian state (meaning, a state characterized by the mean value and the variance of the signal inside it). The time-series signals are sliced in consecutive windows of duration $\Delta t$, and the windows close enough to the state's mean are classified as belonging to that state. These signals are then removed from the analysis, in order to enhance the resolution on the still unclassified signals at the next iteration. At the end of the process each signal windows is thus either classified in one of the identified states, or labeled as ``unclassified" at that specific time resolution. To discard statistically irrelevant clusters, populations below 1\% in this study are removed from the classified data fractions.

Performing this analysis at different values of the time resolution $\Delta t$ allows then to automatically identify the optimal choice of $\Delta t$ that maximizes the number of environments correctly separated, and minimizes the fraction of unclassified points. Complete details can be found in~\cite{becchi_layer-by-layer_2024}. Onion Clustering was performed using the Dynsight~\cite{noauthor_dynsight_nodate} package. 
}

\small{
\subsection{The descriptors' ``evaluation space"}
\label{subs:eval_space}
Principal components analysis was computed using the Scikit-learn python package~\cite{pedregosa_scikit-learn_2011}. Distances between PCA scores are finally classified using SciPy euclidean hierarchical clustering tool~\cite{bar-joseph_fast_2001, mullner_modern_2011}.
}

\section*{Data availability statement}
The TIP4P/ICE simulation trajectory used in this study is available at~\cite{zenodo_water}. 

\noindent The code for performing all the analyses reported in this paper is openly available at the following URL: \href{https://github.com/GMPavanLab/Descriptors.git}{https://github.com/GMPavanLab/Descriptors.git}. 

\section*{Acknowledgments}
G.M.P. acknowledges the funding received by the European Research Council under the European Union’s Horizon 2020 research and innovation program (grant agreement no. 818776–DYNAPOL).

\printbibliography

@online{zenodo_water,
  author={Caruso, Cristina and Cardellini, Annalisa and Crippa, Martina and Rapetti, Daniele and Pavan, Giovanni M.},
  title={Research data supporting: "TimeSOAP: Tracking high-dimensional fluctuations in complex molecular systems via time variations of SOAP spectra"},
  year={2023},
  url={https://doi.org/10.5281/zenodo.7962819},
  note={Accessed: 2024-07-15}
}

@article{gardin_classifying_2022,
	title = {Classifying soft self-assembled materials via unsupervised machine learning of defects},
	volume = {5},
	doi = {10.1038/s42004-022-00699-z},
	pages = {1--15},
	number = {1},
	journaltitle = {Communications Chemistry},
	author = {Gardin, Andrea and Perego, Claudio and Doni, Giovanni and Pavan, Giovanni M.},
	year = {2022},
}

@article{capelli_data-driven_2021,
	title = {A Data-Driven Dimensionality Reduction Approach to Compare and Classify Lipid Force Fields},
	volume = {125},
	doi = {10.1021/acs.jpcb.1c02503},
	pages = {7785--7796},
	number = {28},
	journaltitle = {The Journal of Physical Chemistry. B},
	author = {Capelli, Riccardo and Gardin, Andrea and Empereur-Mot, Charly and Doni, Giovanni and Pavan, Giovanni M.},
	year = {2021},
}

@article{giovambattista_structural_2005,
	title = {Structural order in glassy water},
	volume = {71},
	doi = {10.1103/PhysRevE.71.061505},
	pages = {061505},
	number = {6},
	journaltitle = {Physical Review E},
	author = {Giovambattista, Nicolas and Debenedetti, Pablo G. and Sciortino, Francesco and Stanley, H. Eugene},
	year = {2005},
}

@article{chau_new_1998,
	title = {A new order parameter for tetrahedral configurations},
	volume = {93},
	doi = {10.1080/002689798169195},
	pages = {511--518},
	number = {3},
	journaltitle = {Molecular Physics},
	author = {Chau, P.-L. and Hardwick, A. J.},
	year = {1998},
}

@article{cuthbertson_mixturelike_2011,
	title = {Mixturelike Behavior Near a Liquid-Liquid Phase Transition in Simulations of Supercooled Water},
	volume = {106},
	doi = {10.1103/PhysRevLett.106.115706},
	pages = {115706},
	number = {11},
	journaltitle = {Physical Review Letters},
	author = {Cuthbertson, Megan J. and Poole, Peter H.},
	year = {2011},
}

@article{pedregosa_scikit-learn_2011,
	title = {Scikit-learn: Machine Learning in Python},
	volume = {12},
	pages = {2825--2830},
	number = {85},
	journaltitle = {Journal of Machine Learning Research},
	author = {Pedregosa, Fabian and Varoquaux, Gaël and Gramfort, Alexandre and Michel, Vincent and Thirion, Bertrand and Grisel, Olivier and Blondel, Mathieu and Prettenhofer, Peter and Weiss, Ron and Dubourg, Vincent and Vanderplas, Jake and Passos, Alexandre and Cournapeau, David and Brucher, Matthieu and Perrot, Matthieu and Duchesnay, Édouard},
	year = {2011},
}

@article{himanen_dscribe_2020,
	title = {{DScribe}: Library of descriptors for machine learning in materials science},
	volume = {247},
	doi = {10.1016/j.cpc.2019.106949},
	pages = {106949},
	journaltitle = {Computer Physics Communications},
	author = {Himanen, Lauri and Jäger, Marc O. J. and Morooka, Eiaki V. and Federici Canova, Filippo and Ranawat, Yashasvi S. and Gao, David Z. and Rinke, Patrick and Foster, Adam S.},
	year = {2020},
}

@article{laakso_updates_2023,
	title = {Updates to the {DScribe} library: New descriptors and derivatives},
	volume = {158},
	doi = {10.1063/5.0151031},
	pages = {234802},
	number = {23},
	journaltitle = {The Journal of Chemical Physics},
	author = {Laakso, Jarno and Himanen, Lauri and Homm, Henrietta and Morooka, Eiaki V. and Jäger, Marc O. J. and Todorović, Milica and Rinke, Patrick},
	year = {2023},
}

@article{abraham_gromacs_2015,
	title = {{GROMACS}: High performance molecular simulations through multi-level parallelism from laptops to supercomputers},
	volume = {1-2},
	doi = {10.1016/j.softx.2015.06.001},
	pages = {19--25},
	journaltitle = {{SoftwareX}},
	author = {Abraham, Mark James and Murtola, Teemu and Schulz, Roland and Páll, Szilárd and Smith, Jeremy C. and Hess, Berk and Lindahl, Erik},
	year = {2015},
}

@article{nagy_hierarchical_2010,
	title = {Hierarchical group dynamics in pigeon flocks},
	volume = {464},
	doi = {10.1038/nature08891},
	pages = {890--893},
	number = {7290},
	journaltitle = {Nature},
	author = {Nagy, Máté and Ákos, Zsuzsa and Biro, Dora and Vicsek, Tamás},
	year = {2010},
}

@article{bar-joseph_fast_2001,
	title = {Fast optimal leaf ordering for hierarchical clustering},
	volume = {17},
	doi = {10.1093/bioinformatics/17.suppl_1.S22},
	pages = {S22--S29},
	journaltitle = {Bioinformatics},
	author = {Bar-Joseph, Ziv and Gifford, David K. and Jaakkola, Tommi S.},
	year = {2001},
}

@misc{mullner_modern_2011,
      title={Modern hierarchical, agglomerative clustering algorithms}, 
      author={Daniel Müllner},
      year={2011},
      eprint={1109.2378},
      archivePrefix={arXiv},
      primaryClass={stat.ML},
      url={https://arxiv.org/abs/1109.2378}, 
}

@article{harrington_liquid-liquid_1997,
	title = {Liquid-Liquid Phase Transition: Evidence from Simulations},
	volume = {78},
	doi = {10.1103/PhysRevLett.78.2409},
	pages = {2409--2412},
	number = {12},
	journaltitle = {Physical Review Letters},
	author = {Harrington, Stephen and Zhang, Rong and Poole, Peter H. and Sciortino, Francesco and Stanley, H. Eugene},
	year = {1997},
}

@article{aminikhanghahi_survey_2017,
	title = {A survey of methods for time series change point detection},
	volume = {51},
	doi = {10.1007/s10115-016-0987-z},
	pages = {339--367},
	number = {2},
	journaltitle = {Knowledge and Information Systems},
	author = {Aminikhanghahi, Samaneh and Cook, Diane J.},
	year = {2017},
}

@article{aghabozorgi_time-series_2015,
	title = {Time-series clustering – A decade review},
	volume = {53},
	doi = {10.1016/j.is.2015.04.007},
	pages = {16--38},
	journaltitle = {Information Systems},
	author = {Aghabozorgi, Saeed and Seyed Shirkhorshidi, Ali and Ying Wah, Teh},
	year = {2015},
}

@article{shiraj_anomaly_2024,
	title = {Anomaly detection in financial time series data via mapper algorithm and {DBSCAN} clustering},
	volume = {13},
	doi = {10.30574/wjaets.2024.13.1.0396},
	pages = {070--084},
	number = {1},
	journaltitle = {World Journal of Advanced Engineering Technology and Sciences},
	author = {Shiraj, Md Morshed Bin and Rahman, Md Mizanur and Al-Imran, Md and Liza, Mst Zinia Afroz and Murshed, Md Masum and Akhter, Nasima},
	year = {2024},
}

@article{nayar_comparison_2011,
	title = {Comparison of Tetrahedral Order, Liquid State Anomalies, and Hydration Behavior of {mTIP}3P and {TIP}4P Water Models},
	volume = {7},
	doi = {10.1021/ct2002732},
	pages = {3354--3367},
	number = {10},
	journaltitle = {Journal of Chemical Theory and Computation},
	author = {Nayar, Divya and Agarwal, Manish and Chakravarty, Charusita},
	year = {2011},
}

@inbook{kathirgamanathan_feature_2021,
   title={A Feature Selection Method for Multi-dimension Time-Series Data},
   ISBN={9783030657420},
   ISSN={1611-3349},
   url={http://dx.doi.org/10.1007/978-3-030-65742-0_15},
   DOI={10.1007/978-3-030-65742-0_15},
   booktitle={Advanced Analytics and Learning on Temporal Data},
   publisher={Springer International Publishing},
   author={Kathirgamanathan, Bahavathy and Cunningham, Pádraig},
   year={2020},
   pages={220–231} }

@article{abascal_potential_2005,
	title = {A potential model for the study of ices and amorphous water: {TIP}4P/Ice},
	volume = {122},
	doi = {10.1063/1.1931662},
	pages = {234511},
	number = {23},
	journaltitle = {The Journal of Chemical Physics},
	author = {Abascal, J. L. F. and Sanz, E. and García Fernández, R. and Vega, C.},
	year = {2005},
}

@misc{perrone_unsupervised_2024,
      title={Unsupervised Tracking of Local and Collective Defects Dynamics in Metals Under Deformation}, 
      author={Mattia Perrone and Matteo Cioni and Massimo Delle Piane and Giovanni Maria Pavan},
      year={2024},
      eprint={2410.20999},
      archivePrefix={arXiv},
      primaryClass={cond-mat.mtrl-sci},
      url={https://arxiv.org/abs/2410.20999}, 
}

@misc{caruso_classification_2024,
      title={Classification and Spatiotemporal Correlation of Dominant Fluctuations in Complex Dynamical Systems}, 
      author={Cristina Caruso and Martina Crippa and Annalisa Cardellini and Matteo Cioni and Mattia Perrone and Massimo Delle Piane and Giovanni M. Pavan},
      year={2024},
      eprint={2409.18844},
      archivePrefix={arXiv},
      primaryClass={physics.chem-ph},
      url={https://arxiv.org/abs/2409.18844}, 
}

@article{donkor_beyond_2024,
	title = {Beyond Local Structures in Critical Supercooled Water through Unsupervised Learning},
	volume = {15},
	doi = {10.1021/acs.jpclett.4c00383},
	pages = {3996--4005},
	number = {15},
	journaltitle = {The Journal of Physical Chemistry Letters},
	author = {Donkor, Edward Danquah and Offei-Danso, Adu and Rodriguez, Alex and Sciortino, Francesco and Hassanali, Ali},
	year = {2024},
}

@article{becchi_layer-by-layer_2024,
	title = {Layer-by-layer unsupervised clustering of statistically relevant fluctuations in noisy time-series data of complex dynamical systems},
	volume = {121},
	doi = {10.1073/pnas.2403771121},
	pages = {e2403771121},
	number = {33},
	journaltitle = {Proceedings of the National Academy of Sciences},
	author = {Becchi, Matteo and Fantolino, Federico and Pavan, Giovanni M.},
	year = {2024},
}

@article{butler_change_2024,
	title = {Change point detection of events in molecular simulations using dupin},
	volume = {304},
	doi = {10.1016/j.cpc.2024.109297},
	pages = {109297},
	journaltitle = {Computer Physics Communications},
	author = {Butler, Brandon L. and Fijan, Domagoj and Glotzer, Sharon C.},
	year = {2024},
}

@article{crippa_machine_2023,
	title = {Machine learning of microscopic structure-dynamics relationships in complex molecular systems},
	volume = {4},
	doi = {10.1088/2632-2153/ad0fa5},
	pages = {045044},
	number = {4},
	journaltitle = {Machine Learning: Science and Technology},
	author = {Crippa, Martina and Cardellini, Annalisa and Cioni, Matteo and Csányi, Gábor and Pavan, Giovanni M.},
	year = {2023},
}

@article{caruso_timesoap_2023,
	title = {{TimeSOAP}: Tracking high-dimensional fluctuations in complex molecular systems via time variations of {SOAP} spectra},
	volume = {158},
	doi = {10.1063/5.0147025},
	pages = {214302},
	number = {21},
	journaltitle = {The Journal of Chemical Physics},
	author = {Caruso, Cristina and Cardellini, Annalisa and Crippa, Martina and Rapetti, Daniele and Pavan, Giovanni M.},
	year = {2023},
}

@article{crippa_detecting_2023,
	title = {Detecting dynamic domains and local fluctuations in complex molecular systems via timelapse neighbors shuffling},
	volume = {120},
	doi = {10.1073/pnas.2300565120},
	pages = {e2300565120},
	number = {30},
	journaltitle = {Proceedings of the National Academy of Sciences},
	author = {Crippa, Martina and Cardellini, Annalisa and Caruso, Cristina and Pavan, Giovanni M.},
	year = {2023},
}

@article{nigam_recursive_2020,
	title = {Recursive evaluation and iterative contraction of N-body equivariant features},
	volume = {153},
	doi = {10.1063/5.0021116},
	pages = {121101},
	number = {12},
	journaltitle = {The Journal of Chemical Physics},
	author = {Nigam, Jigyasa and Pozdnyakov, Sergey and Ceriotti, Michele},
	year = {2020},
}

@article{drautz_atomic_2019,
	title = {Atomic cluster expansion for accurate and transferable interatomic potentials},
	volume = {99},
	doi = {10.1103/PhysRevB.99.014104},
	pages = {014104},
	number = {1},
	journaltitle = {Physical Review B},
	author = {Drautz, Ralf},
	year = {2019},
}

@article{bartok_representing_2013,
	title = {On representing chemical environments},
	volume = {87},
	doi = {10.1103/PhysRevB.87.184115},
	pages = {184115},
	number = {18},
	journaltitle = {Physical Review B},
	author = {Bartók, Albert P. and Kondor, Risi and Csányi, Gábor},
	year = {2013},
}

@article{donkor_machine-learning_2023,
	title = {Do Machine-Learning Atomic Descriptors and Order Parameters Tell the Same Story? The Case of Liquid Water},
	volume = {19},
	doi = {10.1021/acs.jctc.2c01205},
	pages = {4596--4605},
	number = {14},
	journaltitle = {Journal of Chemical Theory and Computation},
	author = {Donkor, Edward Danquah and Laio, Alessandro and Hassanali, Ali},
	year = {2023},
}

@article{uhrin_through_2021,
	title = {Through the eyes of a descriptor: Constructing complete, invertible descriptions of atomic environments},
	volume = {104},
	doi = {10.1103/PhysRevB.104.144110},
	pages = {144110},
	number = {14},
	journaltitle = {Physical Review B},
	author = {Uhrin, Martin},
	year = {2021},
}

@article{musil_physics-inspired_2021,
	title = {Physics-Inspired Structural Representations for Molecules and Materials},
	volume = {121},
	doi = {10.1021/acs.chemrev.1c00021},
	pages = {9759--9815},
	number = {16},
	journaltitle = {Chemical Reviews},
	author = {Musil, Felix and Grisafi, Andrea and Bartók, Albert P. and Ortner, Christoph and Csányi, Gábor and Ceriotti, Michele},
	year = {2021},
}

@book{schmidt_human-based_2023,
	title = {Human-Based and Automatic Feature Ideation for Time Series Data: A Comparative Study},
	isbn = {978-3-03868-222-6},
	url = {10.2312/eurova.20231089},
	publisher = {The Eurographics Association},
	author = {Schmidt, Johanna and Piringer, Harald and Mühlbacher, Thomas and Bernard, Jürgen},
	year = {2023},
}

@article{cioni_innate_2023,
	title = {Innate dynamics and identity crisis of a metal surface unveiled by machine learning of atomic environments},
	volume = {158},
	doi = {10.1063/5.0139010},
	pages = {124701},
	number = {12},
	journaltitle = {The Journal of Chemical Physics},
	author = {Cioni, Matteo and Polino, Daniela and Rapetti, Daniele and Pesce, Luca and Delle Piane, Massimo and Pavan, Giovanni M.},
	year = {2023},
}

@article{de_marco_controlling_2021,
	title = {Controlling Exchange Pathways in Dynamic Supramolecular Polymers by Controlling Defects},
	volume = {15},
	doi = {10.1021/acsnano.1c01398},
	pages = {14229--14241},
	number = {9},
	journaltitle = {{ACS} Nano},
	author = {de Marco, Anna L. and Bochicchio, Davide and Gardin, Andrea and Doni, Giovanni and Pavan, Giovanni M.},
	year = {2021},
}

@article{bochicchio_how_2019,
	title = {How Defects Control the Out-of-Equilibrium Dissipative Evolution of a Supramolecular Tubule},
	volume = {13},
	doi = {10.1021/acsnano.8b09523},
	pages = {4322--4334},
	number = {4},
	journaltitle = {{ACS} Nano},
	author = {Bochicchio, Davide and Kwangmettatam, Supaporn and Kudernac, Tibor and Pavan, Giovanni M.},
	year = {2019},
}

@article{gasparotto_identifying_2020,
	title = {Identifying and Tracking Defects in Dynamic Supramolecular Polymers},
	volume = {124},
	doi = {10.1021/acs.jpcb.9b11015},
	pages = {589--599},
	number = {3},
	journaltitle = {The Journal of Physical Chemistry B},
	author = {Gasparotto, Piero and Bochicchio, Davide and Ceriotti, Michele and Pavan, Giovanni M.},
	year = {2020},
}

@article{baletto_structural_2019,
	title = {Structural properties of sub-nanometer metallic clusters},
	volume = {31},
	doi = {10.1088/1361-648X/aaf989},
	pages = {113001},
	number = {11},
	journaltitle = {Journal of Physics: Condensed Matter},
	author = {Baletto, Francesca},
	year = {2019},
}

@article{wolde_enhancement_1997,
	title = {Enhancement of Protein Crystal Nucleation by Critical Density Fluctuations},
	volume = {277},
	doi = {10.1126/science.277.5334.1975},
	pages = {1975--1978},
	number = {5334},
	journaltitle = {Science},
	author = {Wolde, Pieter Rein ten and Frenkel, Daan},
	year = {1997},
}

@software{noauthor_dynsight_nodate,
	title = {dynsight: Simplifies analysis of Molecular Dynamics simulations.},
	shorttitle = {dynsight},
	version = {2024.10.10},
        url = {https://github.com/GMPavanLab/dynsight}
}

@book{mantegna1999introduction,
  title={Introduction to econophysics: correlations and complexity in finance},
  author={Mantegna, Rosario N and Stanley, H Eugene},
  year={1999},
  publisher={Cambridge university press}
}

@inproceedings{keogh2002finding,
  title={Finding surprising patterns in a time series database in linear time and space},
  author={Keogh, Eamonn and Lonardi, Stefano and Chiu, Bill'Yuan-chi'},
  booktitle={Proceedings of the eighth ACM SIGKDD international conference on Knowledge discovery and data mining},
  pages={550--556},
  year={2002},
  doi={10.1145/775047.775128},
  editors={Zaiane, Osmar R. and Goebel, Randy},
}

@article{gupta2013outlier,
  title={Outlier detection for temporal data: A survey},
  author={Gupta, Manish and Gao, Jing and Aggarwal, Charu C and Han, Jiawei},
  journal={IEEE Transactions on Knowledge and data Engineering},
  volume={26},
  number={9},
  pages={2250--2267},
  year={2013},
  publisher={IEEE},
  doi={10.1109/TKDE.2013.184},
}

@misc{lionello2024relevant,
      title={Relevant, hidden, and frustrated information in high-dimensional analyses of complex dynamical systems with internal noise}, 
      author={Chiara Lionello and Matteo Becchi and Simone Martino and Giovanni M. Pavan},
      year={2024},
      eprint={2412.09412},
      archivePrefix={arXiv},
      primaryClass={physics.chem-ph},
      url={https://arxiv.org/abs/2412.09412}, 
}

\newpage
\setcounter{section}{0}
\setcounter{subsection}{0}
\setcounter{figure}{0}
\renewcommand{\thefigure}{S\arabic{figure}}

\section*{Supporting Information}
\subsection*{Supporting text}
\label{SI:text}
\subsubsection*{Evaluation dataset: computation details}
The evaluation dataset is built computing the sequent quantities:
\begin{itemize}
    \item \textbf{State mode}
    \begin{itemize}
        \item Indicates the most frequently number of states detected for each $\Delta t$ taken into account.
    \end{itemize}
    \item \textbf{Counts}
    \begin{itemize}
        \item Provide the counts of occurrences for each possible state in the dataset (0 to 4: ice, water, interface, double interface).
    \end{itemize}
    \item \textbf{Information lost fraction}
    \begin{itemize}
        \item The average value of the information lost fraction for each $\Delta t$ taken into account.
        \item The standard deviation value of the information lost fraction for each $\Delta t$ taken into account.
    \end{itemize}
    \item \textbf{$\chi$ parameter}
    \begin{itemize}
        \item The average value of the $\chi$ parameter for each $\Delta t$ taken into account.
        \item The standard deviation value of the $\chi$ parameter for each $\Delta t$ taken into account.
    \end{itemize}
    \item \textbf{Time resolution threshold}
    \begin{itemize}
        \item The $\Delta t$ at which the information lost reach or goes over the 50\%
    \end{itemize}
\end{itemize}
Data belonging to classifications that resulted in more than 50 \% of lost information were not considered.

\newpage
\subsection*{Supporting Figures}

\begin{figure}[H]
    \centering
    \includegraphics[width=\textwidth]{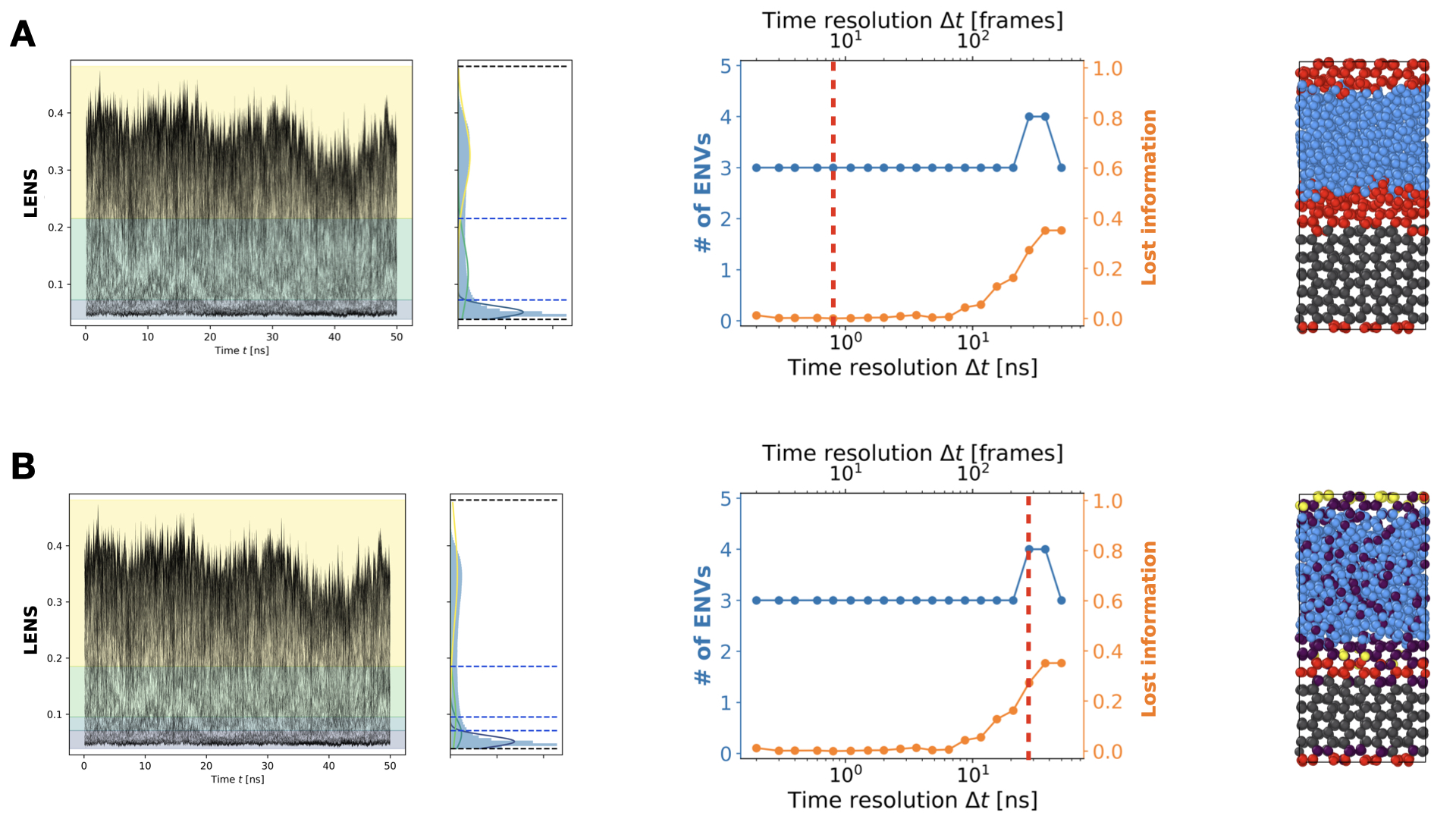}
        \caption{{\bf Denoised LENS}. Onion Clustering analysis on denoised LENS descriptor, for $\Delta t = 0.8$ ns (\textbf{A}) and $\Delta t = 27.8$ ns (\textbf{B})}
    \label{figS1}
\end{figure}

\begin{figure}[H]
    \centering
    \begin{subfigure}[b]{0.45\textwidth}
        \centering
        \includegraphics[width=\textwidth]{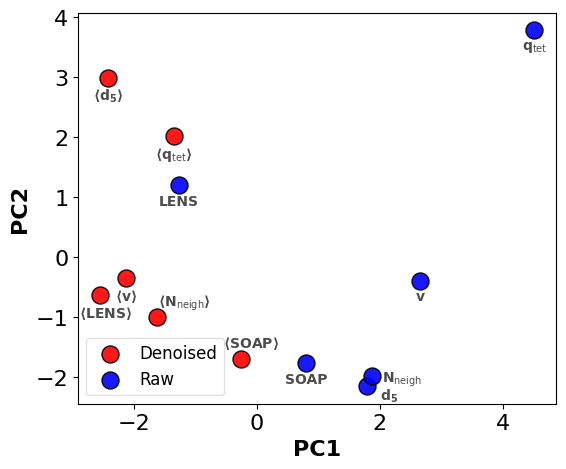}
        \label{pc1pc3}
    \end{subfigure}%
    \hfill
    \begin{subfigure}[b]{0.45\textwidth}
        \centering
        \includegraphics[width=\textwidth]{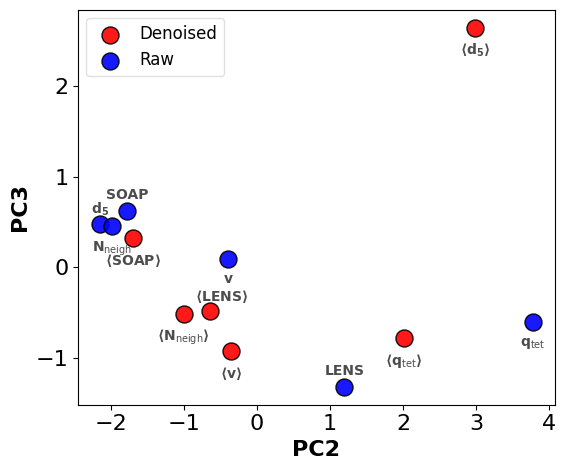}
        \label{pc2pc3}
    \end{subfigure}
    \caption{{\bf Evaluation space: other PCs} To further illustrate the analysis presented in the manuscript, we provide additional PCA views here to offer a more comprehensive overview of the data distribution across the first three principal components. PC1 vs PC2 (left) and PC2 vs PC3 (right) principal component analysis scores.}
    \label{figS2}
\end{figure}

\end{document}